\documentstyle[12pt,graphicx]{article}                
\begin{document}
%This is dvips(k) 5.86 Cobegin{document}
\baselineskip 18pt
%t
\def\today{\ifcase\month\or
 January\or February\or March\or April\or May\or June\or
 July\or August\or September\or October\or November\or December\fi
 \space\number\day, \number\year}
%spacenumberday,: Command not found

%
\def\thebibliography#1{\section*{References\markboth
 {References}{References}}\list
 {[\arabic{enumi}]}{\settowidth\labelwidth{[#1]}
 \leftmargin\labelwidth
 \advance\leftmargin\labelsep
 \usecounter{enumi}}
 \def\newblock{\hskip .11em plus .33em minus .07em}
 \sloppy
 \sfcode`\.=1000\relax}
\let\endthebibliography=\endlist
\def\beq{\begin{equation}}
\def\eeq{\end{equation}}
\def\beqn{\begin{eqnarray}}
\def\eeqn{\end{eqnarray}}
\def\rmuu{\gamma^{\mu}}
\def\rmud{\gamma_{\mu}}
\def\PL{{1-\gamma_5\over 2}}
\def\PR{{1+\gamma_5\over 2}}
\def\sinW2{\sin^2\theta_W}
\def\AEM{\alpha_{EM}}
\def\mul{M_{\tilde{u} L}^2}
\def\mur{M_{\tilde{u} R}^2}
\def\mdl{M_{\tilde{d} L}^2}
\def\mdr{M_{\tilde{d} R}^2}
\def\mz2{M_{z}^2}
\def\c2b{\cos 2\beta}
\def\au{A_u}         
\def\ad{A_d}
\def\cob{\cot \beta}
\def\v#1{v_#1}
\def\tb{\tan\beta}
\def\epem{$e^+e^-$}
\def\KK{$K^0$-$\overline{K^0}$}
\def\wi{\omega_i}
\def\xj{\chi_j}
\def\Wmu{W_\mu}
\def\Wnu{W_\nu}
\def\m#1{{\tilde m}_#1}
\def\mH{m_H}
\def\mw#1{{\tilde m}_{\omega #1}}
\def\mx#1{{\tilde m}_{\chi^{0}_#1}}
\def\mc#1{{\tilde m}_{\chi^{+}_#1}}
\def\mwi{{\tilde m}_{\omega i}}
\def\mxi{{\tilde m}_{\chi^{0}_i}}
\def\mci{{\tilde m}_{\chi^{+}_i}}
\def\mz{M_z}
\def\sw{\sin\theta_W}
\def\cw{\cos\theta_W}
\def\cb{\cos\beta}
\def\sb{\sin\beta}
\def\rwi{r_{\omega i}}
\def\rxj{r_{\chi j}}
\def\rfp{r_f'}
\def\Kik{K_{ik}}
\def\Fq2{F_{2}(q^2)}
\def\f{\({\cal F}\)}
\def\d1{{\f(\tilde c;\tilde s;\tilde W)+ \f(\tilde c;\tilde \mu;\tilde W)}}
%%%%%%%%%%%%%%%%%%%%%%%%%%%%%%%%%%
\def\tw{\tan\theta_W}
\def\sec2w{sec^2\theta_W}
%%%%%%%%%%%%%%%%%%%%%%%%%%%%%%%%%%

\begin{titlepage}

\begin{center}
{\Large {\bf  { Complete Cubic and Quartic  Couplings of $\bf{16}$ and 
$\overline {\bf{16}}$ in  SO(10) Unification}}}\\
\vskip 0.5 true cm
\vspace{2cm}
\renewcommand{\thefootnote}
{\fnsymbol{footnote}}
 Pran Nath$^{a,b}$ and Raza M. Syed$^b$  
\vskip 0.5 true cm
\end{center}

\noindent
{a. Theoretical Physics Division, CERN CH-1211, Geneve 23, 
Switzerland}\\
{b. Department of Physics, Northeastern University,
Boston, MA 02115-5000, USA\footnote{Permanent address of P.N.}} \\
%\footnote{ $\dagger$ : Permanent address}
\vskip 1.0 true cm
\centerline{\bf Abstract}
\medskip 
\noindent
A recently derived basic theorem on the decomposition of SO(2N) vertices
is used to obtain a complete analytic determination of all SO(10)
invariant cubic superpotential couplings involving $16_{\pm}$ semispinors of 
SO(10) chirality $\pm$ and 
tensor representations. In addition to the superpotential couplings
 computed previously
using the basic theorem involving the 10, 120 and $\overline{126}$ tensor 
representations we compute here couplings involving the 1, 45  
and 210 dimensional tensor representations, i.e., we compute
the $\overline{16}_{\mp}16_{\pm}1$, 
$\overline{16}_{\mp}16_{\pm}45$ and $\overline{16}_{\mp}16_{\pm}210$
Higgs couplings in the superpotential. A complete
determination of dimension five operators in the superpotential  
arising from the mediation of the 1, 45 and 210 dimensional representations 
is also given.
The vector couplings $\overline{16}_{\pm}16_{\pm}1$, 
$\overline{16}_{\pm}16_{\pm}45$ and $\overline{16}_{\pm}16_{\pm}210$
are also analyzed. 
The role of large tensor representations and the possible
application of results derived here in model building are discussed.

\end{titlepage}

\section{Introduction}
The group SO(10) is an interesting possible candidate for unification 
of interactions\cite{georgi} and there has been considerable interest
 recently 
in investigating specific grand unified models based on this group.
Thus SO(10) models have many desirable features allowing for 
all the quarks and leptons of one generation to reside in the 
irreducible 16 plet spinor representation of SO(10) and allowing
for a natural splitting of Higgs doublets and Higgs 
triplets. 
Progress on the explicit computation of SO(10) couplings has been
less dramatic. Thus while good initial progress occured in the
early nineteen eightees in the introduction of oscillator 
techniques\cite{sakita,wilczek,nandi}, there was 
little further progress on this front till recently when
a technique was developed using the oscillator method which allows
for the explicit computation of SO(2N) invariant couplings\cite{ns}.
It was also shown in Ref.\cite{ns} that the new technique is specially
 useful in the
analysis of couplings involving large tensor representations.
Large tensor representations have
already surfaced in several unified models based on SO(10)\cite{gn}
and one needs to address the question of fully evaluating 
couplings involving the 16 plet of matter and Higgs with these tensors.
In Ref.\cite{ns} a complete evaluation of the cubic superpotential 
involving the 16 plet of matter was given. Since
$16\times 16 =10+120_a+126_s$  the evaluations given in 
Ref.\cite{ns} involved $16-16-10$, $16-16-120$ and $16-16-\overline{126}$
couplings. 

In this paper we carry the analysis a step further and 
give a complete evaluation of the $\overline{16}-16$  
couplings which involve the SO(10) tensors 1, 45 and 210.
Further, 
  technically the couplings of $\overline{16}_-16_+$ are not 
  necessarily the same as of $\overline{16}_+16_-$.
  Thus we give a full evaluation of the 
  $\overline{16}_{\mp}16_{\pm}1$,
  $\overline{16}_{\mp}16_{\pm}45$ and the $\overline{16}_{\mp}16_{\pm}120$
  couplings. An analysis of 
  $\overline{16}_{\pm}16_{\pm}1$,$\overline{16}_{\pm}16_{\pm}45$, and
$\overline{16}_{\pm}16_{\pm}210$
 vector couplings is also given.
   The analysis given here will have direct 
  application in the further development of SO(10)
   unified models and in a  fuller understanding  of their 
  detailed structure.
  We wish to point out
  that one may also use purely group theoretic methods to compute
  the Clebsch-Gordon co-efficients  in the expansion
  of SO(10) invariant couplings. Such an approach was used in
  Ref.\cite{anderson} to compute the $E_6$ couplings. Our
  approach is field theoretic  and is  specially suited 
  for the computation of SO(2N) couplings.   The  outline
  of the rest of the paper is as follows: In Sec.2 we give a 
  brief review of the basic theorem derived in Ref.\cite{ns}
  which is central to the computation of SO(2N) invariant
  couplings.
  In Sec.3 we use the basic theorem to compute the 
  superpotential couplings cubic in fields involving 
  $\overline{16}_{\mp}16_{\pm}$ and the 1 and 45 tensor fields.
   In Sec.4 a similar analysis is carried
  out using the 210 multiplet.
  In Sec.5 an analysis is given of the quartic couplings  
  in the superpotential 
  obtained from the elimination of the singlet, the 45 plet and the
   210 plet of heavy Higgs fields from the cubic superpotential. 
   Vector couplings are investigated in Sec.6. 
   In Sec.7 the possible role of 
   large tensor representations in model building is discussed.
   Conclusions are given in Sec.8. Some of the mathematical details
   are discussed in Appendices A and B.

\section{Review of basic theorem for analysis of SO(2N) couplings }
In this section we give a discussion of the oscillator 
method\cite{sakita,wilczek,nandi} 
  together with  a brief discussion of the basic theorem 
  derived in Ref.\cite{ns} which is especially useful in 
   evaluating SO(2N) gauge and Yukawa  couplings involving
   large tensor representations of SO(2N). 
We begin by  defining a set of five fermionic creation 
and annihilation operators $b_i$ and $b_i^{\dagger}$ 
($i=1,...,5$) obeying the anti-commutation rules
\begin{equation}
\{b_i,b_j^{\dagger}\}=\delta_{i}^j;~~~\{b_i,b_j\}=0;
~~~\{b_i^{\dagger},b_j^{\dagger}\}=0
\end{equation}
and represent the set of ten Hermitian operators 
$\Gamma_{\mu}$ ($\mu=1,2,..,10$) by 
\begin{equation}
\Gamma_{2i}= (b_i+ b_i^{\dagger});~~~ 
\Gamma_{2i-1}= -i(b_i- b_i^{\dagger})
\end{equation}
where $\Gamma_{\mu}$ define a rank-10 Clifford algebra,
\begin{equation}
\{\Gamma_{\mu},\Gamma_{\nu}\}=2\delta_{\mu\nu}.
\end{equation}
and $\Sigma_{\mu \nu}=\frac{1}{2i}[\Gamma_\mu,\Gamma_\nu]$
are the 45 generators of SO(10) in the spinor representation.
$\frac{1}{2}(1\pm \Gamma_0)$ where 
$\Gamma_0=i^5  \Gamma_1\Gamma_2...\Gamma_{10}$  are 
the SO(10) chirality operators which 
split the 32-dimensional spinor $\Psi$ into two 
inequivalent spinors through the relation
\begin{equation}
\Psi_{(\pm)}=\frac{1}{2}(1 \pm \Gamma_0)\Psi.
\end{equation}
The semi-spinor $\Psi_{(+)}$ ($\Psi_{(-)}$) 
transforms  as a 16  ($\overline {16}$)  
dimensional irreducible representation of SO(10). $\Psi_{(+)}$ 
($\Psi_{(-)}$) contains  $1+\overline5+10$ 
($1+5+\overline{10}$) in its SU(5) decomposition. 
In terms of their oscillator modes we can expand them as\cite{sakita}
\begin{equation}
|\Psi_{(+)a}>=|0>{\bf M}_a+\frac{1}{2}b_i^{\dagger}b_j^{\dagger}|0>{\bf M}_a^{ij}
+\frac{1}{24}\epsilon^{ijklm}b_j^{\dagger}
b_k^{\dagger}b_l^{\dagger}b_m^{\dagger}|0>{\bf M}_{ai}
\end{equation}
\begin{equation}
|\Psi_{(-)b}>=b_1^{\dagger}b_2^{\dagger}
b_3^{\dagger}b_4^{\dagger}b_5^{\dagger}|0>{\bf N}_b
+\frac{1}{12}\epsilon^{ijklm}b_k^{\dagger}b_l^{\dagger}
b_m^{\dagger}|0>{\bf N}_{bij}+b_i^{\dagger}|0>{\bf N}_b^i
\end{equation}
where the SU(5) singlet state $|0>$ is such that $b_i|0>=0$. The
subscripts $a,b=1,2,3$ are the generation indices.
The group theoretic difference between the chirality $(+)$ and $(-)$ states is
that chirality $(+)$ states are generated by the action of an even number
of creation operators on the vacuum state while the chirality $(-)$ states are 
generated by the action of an odd number of creation operators. The
chirality $(+)$ fields, $|\Psi_{(+)a}>$, $(a=1,2,3)$ can be identified with the
three generations of quarks and leptons, constituted by three sets of $\bar
5+10$ plets of SU(5) and three SU(5) singlet (right handed neutrino) fields. 
The chirality $(-)$ fields, $|\Psi_{(-)}>$ could be Higgs multiplets that
arise in SO(10) model building. The role of the negative chirality fields in
model building will be discussed at the end of Sec.5.
For the sake of completeness we identify the components of a
 16 plet $|\Psi_{(+)a}>$ in terms of particle states 
so that 
\begin{eqnarray}
{\bf M}_a=\nu_{La}^c;~~~~{\bf M}_{a\alpha}=D_{La\alpha}^c;
~~~~{\bf M}_a^{\alpha\beta}=\epsilon^{\alpha\beta\gamma}U_{La\gamma}^c;
~~~~{\bf M}_{a4}=E_{La}^-\nonumber\\
{\bf M}_a^{4\alpha}=U_{La\alpha};
~~~~{\bf M}_{a5}=\nu_{La};~~~~{\bf M}_a^{5\alpha}=D_{La\alpha};
~~~{\bf M}_a^{45}=E_{La}^+
\end{eqnarray}
where $\alpha, \beta, \gamma=1,2,3$ are color indices and we adopt the
convention that all particles are left handed($L$).

Our main focus is the computation of the cubic and quartic couplings
in the superpotential. As already mentioned in the introduction
the couplings of the tensor fields 10, 120 and $\overline{126}$ 
with $16\times 16$ have already been computed in Ref.\cite{ns} and
here we focus on the couplings of the tensor fields 1, 45 and 210
with $\overline{16}\times 16$.  Specifically the interactions of 
  interest in the superpotential involving $16_{\pm}$ semispinors
 are of the form
\begin{equation}
{\mathsf
W}^{(1)}_{-+}=h^{^{(1)}}_{ab}<\widehat{\Psi}_{(-)a}^*|B|\widehat{\Psi}_{(+)b}>\Phi
\end{equation}
\begin{equation}
{\mathsf W}^{(45)}_{-+}=\frac{1}{2!}h^{^{(45)}}_{ab}<\widehat{\Psi}_{(-)a}^*|B
\Sigma_{\mu\nu}|\widehat{\Psi}_{(+)b}>\Phi_{\mu\nu}
\end{equation}
\begin{equation}
{\mathsf W}^{(210)}_{-+}=\frac{1}{4!}
h^{^{(210)}}_{ab}<\widehat{\Psi}_{(-)a}^*|B
\Gamma_{[\mu}\Gamma_{\nu}\Gamma_{\rho}
\Gamma_{\lambda]}
|\widehat{\Psi}_{(+)b}>\Phi_{\mu\nu\rho\lambda}
\end{equation}
where 
\begin{equation}
B=\prod_{\mu =odd}\Gamma_{\mu}= -i\prod_{k=1}^5
(b_k-b_k^{\dagger})
\end{equation}
is an SO(10) charge conjugation operator,
and
\begin{equation}
\Gamma_{[\mu}\Gamma_{\nu}\Gamma_{\rho}
\Gamma_{\lambda]}=\frac{1}{4!}\sum_P(-1)^{\delta_P}
\Gamma_{\mu_{P(1)}}\Gamma_{\nu_{P(2)}}\Gamma_{\rho_{P(3)}}
\Gamma_{\lambda_{P(4)}}
\end{equation}
with $\sum_P$ denoting the sum over all permutations and $\delta_P$
takes on the value $0$ and $1$ for even and odd permutations respectively.
Semi-spinors $\Psi_{(\pm)}$
with a $~\widehat{  }~$ stands for chiral superfields.
The result essential to the analysis of the above SO(2N) 
(N=5) invariant couplings is the theorem \cite{ns} that 
the vertex $\Gamma_{\mu}\Gamma_{\nu}\Gamma_
{\lambda}..\Gamma_{\sigma}$ $\Phi_{\mu\nu\lambda ..\sigma}$ 
where $\Phi_{\mu\nu\lambda ..\sigma}$ could be a large tensor
representation,
can be expanded in the following form 
\begin{eqnarray}
\Gamma_{\mu}\Gamma_{\nu}\Gamma_{\lambda}..\Gamma_{\sigma}
\Phi_{\mu\nu\lambda ...\sigma}= b_i^{\dagger} 
b_j^{\dagger}b_k^{\dagger}...b_n^{\dagger} \Phi_{c_ic_jc_k...c_n}
+\left(b_i b_j^{\dagger}b_k^{\dagger}...b_n^{\dagger} 
\Phi_{\overline c_ic_jc_k...c_n}+~perms\right)\nonumber\\
+\left(b_i b_jb_k^{\dagger}...b_n^{\dagger} \Phi_{\overline c_i\overline
c_jc_{k}...c_n}+~perms\right)+ ...
+\left(b_ib_jb_k...b_{n-1}b_n^{\dagger}\Phi_{\overline c_i\overline c_j
\overline c_k...\overline c_{n-1}c_n}+~perms\right)\nonumber\\
+ b_ib_jb_k...b_n \Phi_{\overline c_i\overline c_j\overline c_k...\overline c_n}
\nonumber\\
\end{eqnarray}
where we have introduced the notation $\Phi_{c_i}=\Phi_{2i}
+i\Phi_{2i-1}$ and $\Phi_{\overline c_i}=\Phi_{2i}-i\Phi_{2i-1}$.
This is extended immediately to define the quantity 
$\Phi_{c_ic_j\overline c_k..}$ with an arbitrary number
of barred and unbarred indices, where each c index can be 
expanded out so that $\Phi_{c_ic_j\overline c_k..}=
\Phi_{2ic_j\overline c_k...}+i\Phi_{2i-1c_j\overline c_k..}$ etc..  Further the
object $\Phi_{c_ic_j\overline c_k...c_n}$ transforms like a reducible
representation of SU(N) which can be further 
decomposed in its irreducible parts.

\section{The 45-plet tensor coupling} 
We first present the result of the trivial 
${\overline {16}\times 16\times 1}$ couplings. Eq.(8) 
 at once gives

\begin{equation}
{\mathsf W}^{(1)}_{-+}=ih^{^{(1)}}_{ab}\left(\widehat {\bf
N}_a^{\bf{T}}\widehat {\bf M}_b-\frac{1}{2}\widehat 
{\bf N}_{aij}^{\bf{T}}\widehat {\bf M}_b^{ij}+\widehat {\bf
N}_a^{i\bf{T}}\widehat
{\bf M}_{bi}\right){\mathsf H}
\end{equation}
where ${\mathsf H}$ is an SO(10) singlet. 
For eg. above  $\widehat {\bf N}_{a}^{\bf{T}}$ represents the
transpose of the chiral superfield, $\widehat {\bf N}_{a}$ etc.. 
A  similar analysis gives
$ {\mathsf W}^{(1)}_{+-}$ and one has                

\begin{eqnarray}
{\mathsf
W}^{(1)}_{+-}=h^{^{(1)}}_{ab}<\widehat{\Psi}_{(+)a}^*|B|\widehat{\Psi}_{(-)b}>
\Phi
~~~~~~~~~~~~~~~~~~~~~~~\nonumber\\   
=ih^{^{(1)}}_{ab}\left(-\widehat {\bf
M}_a^{\bf{T}}\widehat {\bf N}_b+\frac{1}{2}\widehat
{\bf M}_a^{ij\bf{T}}\widehat {\bf N}_{bij}-\widehat {\bf
M}_{ai}^{\bf{T}}\widehat
{\bf N}_b^i\right){\mathsf H}.                 
\end{eqnarray}
\vskip 0.5cm  
\noindent
 To compute the ${\overline {16}\times 16\times 45}$ 
couplings we expand 
the vertex $\Sigma_{\mu\nu}\Phi_{\mu\nu}$ 
 using Eq.(13)  where  $\Phi_{\mu\nu}$ is the 45 plet tensor field
 
\begin{equation}
\Sigma_{\mu\nu}\Phi_{\mu\nu}=\frac{1}{i}\left(b_ib_j
\Phi_{\overline c_i\overline c_j}+b_i^{\dagger}b_j^{\dagger}
\Phi_{c_ic_j}+2b_i^{\dagger}b_j\Phi_{c_i\overline c_j}-
\Phi_{c_n\overline c_n}\right).
\end{equation}
The reducible tensors that enter in the above expansion 
can be decomposed into their irreducible parts as follows

 \begin{eqnarray}
\Phi_{c_n\overline c_n}={\mathsf h};~~~\Phi_{c_i\overline c_j}={\mathsf h}_{j}^i+
\frac{1}{5}\delta_j^i{\mathsf h};~~~
\Phi_{c_ic_j}={\mathsf h}^{ij};~~~\Phi_{\overline c_i\overline c_j}={\mathsf h}_{ij}
\end{eqnarray} 
To normalize the SU(5) Higgs fields contained in the tensor
$\Phi_{\mu\nu}$, we carry out a field redefinition  
\begin{eqnarray}
{\mathsf h}=\sqrt {10}{\mathsf H};~~~{\mathsf h}_{ij}=\sqrt 2{\mathsf H}_{ij};
~~~
{\mathsf h}^{ij}=\sqrt 2{\mathsf H}^{ij};~~~{\mathsf h}_{j}^i=\sqrt 2
{\mathsf H}_{j}^i.
\end{eqnarray}
In terms of the normalized fields the kinetic energy of the 
 45 plet of Higgs\\ 
$-\partial_A\Phi_{\mu\nu}\partial^A\Phi_{\mu\nu}^{\dagger}$ 
takes the form 
\begin{equation}
{\mathsf L}_{kin}^{45-Higgs}=-\partial^A{\mathsf H}\partial_A{\mathsf H}^\dagger
-\frac{1}{2!}
\partial^A{\mathsf H}_{ij}\partial_A{\mathsf H}_{ij}^{\dagger}
-\frac{1}{2!}
\partial^A{\mathsf H}^{ij}\partial_A{\mathsf H}^{ij\dagger}
-\partial_A{\mathsf H}_j^i
\partial^A{\mathsf H}_j^{i\dagger}. 
\end{equation}
The terms in Eq.(19) are only exhibited for the purpose of normalization
and the remaining supersymmetric parts are not
exhibited as their normalizations are rigidly fixed relative to the 
 parts given above\cite{applied}.
Finally, straightforward evaluation of Eq.(9) using Eqs.(16-18) gives
\begin{eqnarray}
{\mathsf W}^{(45)}_{-+}=\frac{1}{\sqrt 2}h^{^{(45)}}_{ab}
[\sqrt {5}\left(\frac{3}{5}\widehat 
{\bf N}_a^{i\bf{T}}\widehat {\bf M}_{bi}+\frac{1}{10}\widehat {\bf
N}^{\bf{T}}_{aij}\widehat {\bf M}_b^{ij}-
\widehat {\bf N}_a^{\bf{T}}\widehat {\bf M}_b\right){\mathsf H}\nonumber\\
+\left(-\widehat
{\bf N}_a^{\bf{T}}\widehat {\bf M}_b^{lm}+\frac{1}{2}\epsilon^{ijklm}
\widehat
{\bf N}^{\bf{T}}_{aij}\widehat {\bf M}_{bk}\right){\mathsf H}_{lm}\nonumber\\
+\left(-\widehat
{\bf N}^{\bf{T}}_{alm}\widehat {\bf M}_b+\frac{1}{2}\epsilon_{ijklm}\widehat
{\bf N}_a^{i\bf{T}}\widehat {\bf M}_b^{jk}\right){\mathsf
H}^{lm}\nonumber\\
+2\left(\widehat {\bf N}^{\bf{T}}_{aik}\widehat {\bf M}_b^{kj}-\widehat
{\bf
N}_a^{j\bf{T}}\widehat {\bf M}_{bi}\right){\mathsf H}_j^i].
\end{eqnarray}
From Eq.(20), one finds that the $\overline {16}_N-16_M-45_H$
couplings  consist of the following SU(5) invariant components:
 $5_N-\overline 5_M-1_H, \overline
{10}_N-10_M-1_H, 1_N-1_M-1_H, 1_N-10_M-\overline {10}_H, \overline
{10}_N-\overline 5_M-\overline {10}_H, \overline {10}_N-1_M-10_H,
5_N-10_M-10_H, \overline {10}_N-10_M-24_H$, and  $5_N-\overline 5_M-24_H$ 
couplings. One can carry out a similar analysis  
for ${\mathsf W}_{+-}^{(45)}$ and one finds

\begin{eqnarray}
{\mathsf                                                             
W}^{(45)}_{+-}=\frac{1}{2!}h^{^{(45)}}_{ab}<\widehat{\Psi}_{(+)a}^*|B
\Sigma_{\mu\nu}|\widehat{\Psi}_{(-)b}>\Phi_{\mu\nu}~~~~~~~~~~~~~~~~~
\nonumber\\                 
=\frac{1}{\sqrt 2}h^{^{(45)}}_{ab}[\sqrt {5}\left(\frac{3}{5}\widehat
{\bf M}_{ai}^{\bf{T}}\widehat {\bf N}_b^i+\frac{1}{10}\widehat {\bf
M}^{ij\bf{T}}_a\widehat {\bf N}_{bij}-
\widehat {\bf M}_a^{\bf{T}}\widehat {\bf N}_b\right){\mathsf
H}\nonumber\\
+\left(-\widehat
{\bf M}_a^{lm\bf{T}}\widehat {\bf N}_b+\frac{1}{2}\epsilon^{ijklm}
\widehat
{\bf M}^{\bf{T}}_{ai}\widehat {\bf N}_{bjk}\right){\mathsf
H}_{lm}\nonumber\\
+\left(-\widehat
{\bf M}^{\bf{T}}_a\widehat {\bf
N}_{blm}+\frac{1}{2}\epsilon_{ijklm}\widehat
{\bf M}_a^{ij\bf{T}}\widehat {\bf N}_b^{k}\right){\mathsf
H}^{lm}\nonumber\\
+2\left(\widehat {\bf M}^{jk\bf{T}}_a\widehat {\bf N}_{bki}-\widehat
{\bf
M}_{ai}^{\bf{T}}\widehat {\bf N}_b^j\right){\mathsf H}_j^i].
\end{eqnarray}

\section{The 210-plet tensor coupling} 
We turn now to the computation of the  
$\overline {16}\times 16\times 210$ couplings. Using
Eq.(13) we decompose the vertex
$\Gamma_{\mu}\Gamma_{\nu}\Gamma_{\rho}\Gamma_{\lambda}
\Phi_{\mu\nu\rho\lambda}$ so that

\begin{eqnarray}
\Gamma_{\mu}\Gamma_{\nu}\Gamma_{\rho}\Gamma_{\lambda}
\Phi_{\mu\nu\rho\lambda}=
4b_i^{\dagger}b_j^{\dagger}b_k^{\dagger}b_l\Phi_{c_ic_jc_k\overline c_l}
+4b_i^{\dagger}b_jb_kb_l\Phi_{c_i\overline c_j\overline c_k\overline c_l}
+b_i^{\dagger}b_j^{\dagger}b_k^{\dagger}
b_l^{\dagger}\Phi_{c_ic_jc_kc_l}\nonumber\\
+b_ib_jb_kb_l\Phi_{\overline c_i
\overline c_j\overline c_k\overline c_l}
-6b_i^{\dagger}b_j^{\dagger}\Phi_{c_ic_jc_m\overline c_m}+
6b_ib_j\Phi_{\overline c_i\overline c_j\overline c_mc_m}\nonumber\\
+3\Phi_{c_m\overline c_mc_n\overline c_n}\nonumber
-12b_i^{\dagger}b_j\Phi_{c_i\overline c_jc_m\overline c_m}
+6b_i^{\dagger}b_j^{\dagger}b_kb_l\Phi_{c_ic_j\overline c_k\overline c_l}.
\nonumber\\
\end{eqnarray}
The tensors that appear above can be decomposed 
into their irreducible parts as follows  
\begin{eqnarray}
\Phi_{c_m\overline c_mc_n\overline c_n}={\mathsf h};~~~\Phi_{\overline c_i\overline c_j
\overline c_k\overline c_l}=\frac{1}{24}\epsilon_{ijklm}{\mathsf h}^m;~~~
\Phi_{c_ic_jc_kc_l}=\frac{1}{24}\epsilon^{ijklm}{\mathsf h}_{m}\nonumber\\
\Phi_{c_ic_jc_m\overline c_m}={\mathsf h}^{ij};~~~\Phi_{\overline c_i\overline c_j
\overline c_mc_m}={\mathsf h}_{ij};~~~\Phi_{c_i\overline
c_jc_m\overline c_m}={\mathsf h}_{j}^i+\frac{1}{5}\delta_j^i{\mathsf h}\nonumber\\
\Phi_{c_ic_j\overline c_k\overline c_l}={\mathsf h}_{kl}^{ij}+\frac{1}{3}
\left(\delta_l^i{\mathsf h}_{k}^j-\delta_k^i{\mathsf h}_{l}^j+
\delta_k^j{\mathsf h}_{l}^i
-\delta_l^j{\mathsf h}_{k}^i\right)+\frac{1}{20}\left(\delta_l^i
\delta_k^j-\delta_k^i\delta_l^j\right){\mathsf h}\nonumber\\
\Phi_{c_ic_jc_k\overline c_l}={\mathsf h}_{l}^{ijk}+\frac{1}{3}
\left(\delta_l^k{\mathsf h}^{ij}-\delta_l^j{\mathsf h}^{ik}
+\delta_l^i{\mathsf h}^{jk}\right)\nonumber\\
\Phi_{\overline c_i\overline c_j\overline c_kc_l}={\mathsf h}_{ijk}^l
+\frac{1}{3}\left(\delta_k^l{\mathsf h}_{ij}-\delta_j^l{\mathsf h}_{ik}+
\delta_i^l{\mathsf h}_{jk}\right)
\end{eqnarray}
where ${\mathsf h}$, ${\mathsf h}^i$, ${\mathsf h}_{i}$, 
${\mathsf h}^{ij}$, ${\mathsf h}_{ij}$, 
${\mathsf h}_{j}^i$, ${\mathsf h}_{l}^{ijk}$; ${\mathsf h}_{jkl}^i$ and 
${\mathsf h}_{kl}^{ij}$ 
are the 1-plet, 5-plet, $\overline 5$-plet, 10-plet, 
$\overline {10}$-plet, 24-plet, 40-plet, $\overline {40}$-plet, and
75-plet representations of SU(5), respectively.  We carry out a field 
redefinition such that 

\begin{eqnarray}
{\mathsf h}=4\sqrt{\frac{5}{3}}{\mathsf H};~~~{\mathsf
h}^i=8\sqrt{6}{\mathsf H}^i;~~~{\mathsf
h}_{i}
=8\sqrt{6}{\mathsf H}_{i}\nonumber\\
{\mathsf h}^{ij}=\sqrt{2}{\mathsf H}^{ij};~~~{\mathsf
h}_{ij}=\sqrt{2}{\mathsf H}_{ij};
~~~{\mathsf h}_{j}^i=\sqrt 2{\mathsf H}_{j}^i\nonumber\\
{\mathsf h}_{l}^{ijk}=\sqrt{\frac{2}{3}}{\mathsf H}_{l}^{ijk};~~~{\mathsf h}_{jkl}^i
=\sqrt{\frac{2}{3}}{\mathsf H}_{jkl}^i;~~~{\mathsf h}_{kl}^{ij}
=\sqrt{\frac{2}{3}}{\mathsf H}_{kl}^{ij}.
\end{eqnarray}

\noindent
Now the kinetic energy for the 210 dimensional Higgs field is
$-\partial_A\Phi_{\mu\nu\rho\lambda}\partial^A
\Phi_{\mu\nu\rho\lambda}^{\dagger}$,
which in terms of the redefined fields takes the form
\begin{eqnarray}
{\mathsf L}_{kin}^{210-Higgs}=-\partial_A{\mathsf H}\partial^A
{\mathsf H}^{\dagger}
-\partial_A{\mathsf H}^i\partial^A{\mathsf
H}^{i\dagger}-\partial_A{\mathsf H}_i\partial^A{\mathsf
H}_{i\dagger}\nonumber\\
-\frac{1}{2!}\partial_A{\mathsf H}^{ij}
\partial^A{\mathsf H}^{ij\dagger}
-\frac{1}{2!}\partial_A{\mathsf H}_{ij}
\partial^A{\mathsf H}_{ij}^{\dagger}  
-\partial_A{\mathsf H}_j^i\partial^A{\mathsf
H}_j^{i\dagger}\nonumber\\
-\frac{1}{3!}\partial_A{\mathsf H}_l^{ijk}
\partial^A{\mathsf H}_l^{ijk\dagger}
-\frac{1}{3!}\partial_A{\mathsf H}^l_{ijk}
\partial^A{\mathsf H}^{l\dagger}_{ijk} 
-\frac{1}{2!}\frac{1}{2!}
\partial_A{\mathsf H}_{kl}^{ij}\partial^A{\mathsf H}_{kl}^{ij\dagger}.
\end{eqnarray}
Evaluation of Eq.(10), using Eq.(22) and the normalization of 
Eq.(24) gives,
\begin{eqnarray}
{\mathsf
W}^{(210)}_{-+}=i\sqrt{\frac{2}{3}}h_{ab}^{^{(210)}}
[\frac{1}{2}\sqrt {\frac{{5}}{2}}\left(
\widehat {\bf N}_a^{\bf{T}}\widehat {\bf M}_b+{\frac{1}{10}}\widehat{\bf
N}_{aij}^{\bf{T}}
\widehat {\bf M}_b^{ij}+\frac{1}{5}\widehat {\bf N}_a^{i\bf{T}}\widehat
{\bf M}_{bi}\right){\mathsf H}\nonumber\\
+\frac{\sqrt 3}{4}\left(\widehat {\bf N}^{\bf{T}}_{alm}\widehat {\bf M}_b
+
\frac{1}{6}\epsilon_{ijklm}\widehat{\bf N}_a^{i\bf{T}}\widehat
{\bf M}_b^{jk}
\right){\mathsf H}^{lm}\nonumber\\           
-\frac{\sqrt{3}}{4}\left(\widehat {\bf N}_a^{\bf{T}}\widehat {\bf M}_b^{lm}
+\frac{1}{6}\epsilon^{ijklm}\widehat{\bf N}^{\bf{T}}_{aij}\widehat
{\bf M}_{bk}
\right){\mathsf H}_{lm}\nonumber\\
-\frac{\sqrt 3}{2}\left(\widehat {\bf N}_a^{j\bf{T}}\widehat {\bf M}_{bi}
+\frac{1}{3}\widehat{\bf N}^{\bf{T}}_{aik}\widehat
{\bf M}_b^{kj}\right){\mathsf H}_j^i\nonumber\\
+\frac{1}{6}\epsilon_{ijklm}\widehat
{\bf N}_a^{i\bf{T}}\widehat {\bf M}_b^{jn}{\mathsf H}_n^{klm}
+\frac{1}{6}\epsilon^{ijklm}\widehat
 {\bf N}^{\bf{T}}_{ain}\widehat {\bf M}_{bj}{\mathsf H}_{klm}^n\nonumber\\
+\frac{1}{4}\widehat
{\bf N}^{\bf{T}}_{aij}\widehat {\bf M}_b^{kl}{\mathsf H}_{kl}^{ij}
+\widehat {\bf N}_a^{\bf{T}}\widehat {\bf M}_{bi}{\mathsf H}^i
+\widehat {\bf N}_a^{i\bf{T}}\widehat {\bf M}_b{\mathsf H}_i].
\end{eqnarray}                                             
We note that $\overline {16}_N-16_M-210_H$
couplings have the SU(5) invariant structure consisting of $1_N-1_M-1_H, 
\overline {10}_N-10_M-1_H, 5_N-\overline 5_M-1_H, \overline
{10}_N-1_M-10_H,\\ 5_N-10_M-10_H, 1_N-10_M-\overline {10}_H, \overline
{10}_N-\overline 5_M-\overline {10}_H, 5_N-\overline 5_M-24_H, \overline
{10}_N-10_M-24_H,\\ 5_N-10_M-40_H, \overline {10}_N-\overline 5_M-\overline
{40}_H, \overline {10}_N-10_M-75_H, 1_N-\overline 5_M-5_H, 5_N-1_M-\overline
5_H$.
 An analysis similar to that for Eq.(26) gives
${\mathsf W}^{(210)}_{+-}$ 
\begin{eqnarray}
{\mathsf
W}^{(210)}_{+-}=\frac{1}{4!}h^{^{(210)}}_{ab}<\widehat{\Psi}_{(+)a}^*|B
\Gamma_{[\mu}\Gamma_{\nu}\Gamma_{\rho}
\Gamma_{\lambda]}|\widehat{\Psi}_{(-)b}>\Phi_{\mu\nu\rho\lambda}~~~~~~~~~~~~~~~~~~~\nonumber\\
=i\sqrt{\frac{2}{3}}h_{ab}^{^{(210)}}
[-\frac{1}{2}\sqrt{\frac{{5}}{2}}\left(
\widehat {\bf M}_a^{\bf{T}}\widehat {\bf
N}_b+{\frac{1}{10}}\widehat{\bf
M}_a^{ij\bf{T}}
\widehat {\bf N}_{bij}+\frac{1}{5}\widehat {\bf M}_{ai}^{\bf{T}}
\widehat {\bf N}_{b}^{i}\right){\mathsf H}\nonumber\\
-\frac{\sqrt{3}}{4}\left(\widehat {\bf M}^{\bf{T}}_{a}\widehat {\bf
N}_{blm}
+\frac{1}{6}\epsilon_{ijklm}\widehat{\bf M}_a^{ij\bf{T}}\widehat
{\bf N}_b^{k}
\right){\mathsf H}^{lm}\nonumber\\
+\frac{\sqrt{3}}{4}\left(\widehat {\bf M}_a^{lm\bf{T}}\widehat {\bf
N}_b
+\frac{1}{6}\epsilon^{ijklm}\widehat{\bf M}^{\bf{T}}_{ai}\widehat
{\bf N}_{bjk}
\right){\mathsf H}_{lm}\nonumber\\
+\frac{\sqrt 3}{2}\left(\widehat {\bf M}_{ai}^{\bf{T}}\widehat {\bf
N}_{b}^j
+\frac{1}{3}\widehat{\bf M}^{jk\bf{T}}_a\widehat
{\bf N}_{bki}\right){\mathsf H}_j^i\nonumber\\
+\frac{1}{12}\epsilon_{ijklm}\widehat
{\bf M}_a^{ij\bf{T}}\widehat {\bf N}_b^{n}{\mathsf H}_n^{klm}
-\frac{1}{12}\epsilon^{ijklm}\widehat
 {\bf M}^{\bf{T}}_{an}\widehat {\bf N}_{bij}{\mathsf
H}_{klm}^n\nonumber\\
-\frac{1}{4}\widehat
{\bf M}^{kl\bf{T}}_{a}\widehat {A\bf N}_{bij}{\mathsf
H}_{kl}^{ij}
-\widehat {\bf M}_{ai}^{\bf{T}}\widehat {\bf N}_{b}{\mathsf
H}^i
-\widehat {\bf M}_{a}^{\bf{T}}\widehat {\bf N}_b^i{\mathsf H}_i].
\end{eqnarray} 
We note that the couplings of ${\mathsf W}^{(210)}_{-+}$ 
are in general not the same as in ${\mathsf W}^{(210)}_{+-}$  .
Thus some of the terms have signs which are opposite in the
two sets. Further, we note that there are in general two ways
in which the 40 plet and the $\overline {40}$ plet can contract
with the matter fields. For the case of ${\mathsf W}^{(210)}_{-+}$ 
one of the 40 plet tensor index contracts with the tensor index
of the 10 plet of matter and similarly one  of the tensor index
on the $\overline {40}$ contracts with the tensor index in the
$\bar{10}$  of the $\overline{16}$ (see Eq.(26)). 
However, in the ${\mathsf W}^{(210)}_{+-}$
couplings this is not the case. Here 
one of the tensor index of 40 plet contracts with the tensor index
in of the 5 plet of matter and similarly one  of the tensor index
in $\overline {40}$ contracts with the tensor index in the
$\bar{5}$ plet of matter (see Eq.27)).

\section{Quartic Couplings of the form
$\overline {\bf{16}}~\bf{16}~\overline {\bf{16}}~\bf{16}$}
In phenomenological analyses one generally needs more than one Higgs 
representations. Hence to keep the analysis very general we not only keep
the generational indices but also allow for mixing among Higgs
representations. To that end, we assume several Higgs representations of the 
same kind: $\Phi_{\cal X}, \Phi_{\mu\nu {\cal Y}}, \Phi_{\mu\nu\rho\lambda
{\cal Z}}$.
Consider the superpotential
\begin{equation}
{\mathsf W}^{^{(16 \times \overline {16})}}=
{\mathsf W}_{Higgs}^{^{(16 \times \overline {16})}}
+{\mathsf W}_{mass}^{^{(16 \times \overline {16})}}
\end{equation}
where
\begin{equation}
{\mathsf W}_{Higgs}^{^{(16 \times \overline {16})}}
={\mathsf W}^{(1)'}_{-+}+{\mathsf W}^{(45)'}_{-+}+{\mathsf W}^{(210)'}_{-+}
\end{equation}  
and
\begin{equation}
{\mathsf W}_{mass}^{^{(16 \times \overline {16})}}
=\frac{1}{2}\Phi_{\cal X}{\cal M}^{^{(1)}}_{{\cal X}{\cal X}'}\Phi_{{\cal
X}'}
+\frac{1}{2}\Phi_{\mu\nu {\cal Y}} {\cal M}^{^{(45)}}_{{\cal Y}{\cal Y}'}
\Phi_{\mu\nu {\cal Y}'}  
+\frac{1}{2}\Phi_{\mu\nu\rho\lambda {\cal Z}} {\cal M}^{^{(210)}}
_{{\cal Z}{\cal Z}'}
\Phi_{\mu\nu\rho\lambda {\cal Z}'}.  
\end{equation}
The terms ${\mathsf W}^{(1)'}_{-+}$, ${\mathsf W}^{(45)'}_{-+}$, 
${\mathsf W}^{(210)'}_{-+}$                                                   
in Eq.(29) are the same as those given by Eqs.(8), (9), and (10) except
that the tensors $\Phi$, $\Phi_{\mu\nu}$, and $\Phi_{\mu\nu\rho\lambda}$
are replaced by $f_{_{{\cal X}}}^{^{(1)}}\Phi_{\cal X}$, $f_{_{{\cal Y}}}^{^{(45)}}
\Phi_{\mu\nu {\cal Y}}$, and $f_{_{{\cal Z}}}^{^{(210)}} 
\Phi_{\mu\nu\rho\lambda {\cal Z}}$,
respectively. 
We next eliminate $\Phi_{\cal X}, \Phi_{\mu\nu {\cal Y}},
\Phi_{\mu\nu\rho\lambda {\cal Z}}$ as
superheavy dimension-5 operators using the F-flatness conditions:
\begin{equation}
\frac{\partial {\mathsf W}^{^{(16 \times \overline {16})}}}{\partial
\Phi_{\cal X}}=0;~~~ 
\frac{\partial {\mathsf W}^{^{(16 \times \overline {16})}}}{\partial
\Phi_{\mu\nu {\cal Y}}}=0;~~~
\frac{\partial {\mathsf W}^{^{(16 \times \overline {16})}}}{\partial
\Phi_{\mu\nu\rho\lambda {\cal Z}}}=0.
\end{equation} 
The above leads to
\begin{equation}
{\mathsf W}^{(\overline {16} \times 16)}_{dim-5} ={\cal I}_{1}+{\cal I}_{45}+{\cal
I}_{210}.
\end{equation}
${\cal I}_{1}$, ${\cal I}_{45}$ and ${\cal I}_{210}$
can be computed quite straightforwardly by integrating out the 
heavy SO(10) singlet, 45 and 210 plets fields in the superpotential. 
Details are given in Appendix A. We record here the results.
\begin{eqnarray}
{\cal I}_{1}=\frac{1}{2}\lambda_{ab,cd}^{^{(1)}}
[-\widehat{\bf N}_{aij}^{\bf{T}}\widehat {\bf M}^{ij}_{b}\widehat{\bf
N}_{ckl}^{\bf{T}}\widehat 
{\bf M}^{kl}_d
+4\widehat{\bf N}^{i\bf{T}}_a\widehat {\bf M}_{bi}\widehat{\bf
N}_{cjk}^{\bf{T}}\widehat 
{\bf M}^{jk}_d
-4\widehat{\bf N}^{i\bf{T}}_a\widehat {\bf M}_{bi}\widehat{\bf N}^{j\bf{T}}_c 
\widehat {\bf M}_{dj}\nonumber\\
+4\widehat{\bf N}^{\bf{T}}_a\widehat {\bf M}_b\widehat{\bf
N}^{\bf{T}}_{cij}\widehat
{\bf M}^{ij}_d
-8\widehat{\bf N}^{\bf{T}}_a\widehat {\bf M}_b\widehat{\bf
N}^{i\bf{T}}_c\widehat
{\bf M}_{di}
-4\widehat{\bf N}_a^{\bf{T}}\widehat {\bf M}_b\widehat{\bf
N}_c^{\bf{T}}\widehat {\bf M}_d]
\end{eqnarray}
\begin{eqnarray}
{\cal I}_{45}=\left(-4\lambda_{ad,cb}^{^{(45)}}+11\lambda_{ab,cd}^{^{(45)}}
\right)\widehat{\bf N}^{i\bf{T}}_a\widehat {\bf M}_{bi}\widehat{\bf
N}^{j\bf{T}}_c\widehat {\bf M}_{dj}
+8\left(\lambda_{ad,cb}^{^{(45)}}+\lambda_{ab,cd}^{^{(45)}}
\right)\widehat{\bf N}^{i\bf{T}}_a\widehat {\bf M}_{bj}\widehat{\bf
N}_{cik}^{\bf{T}}\widehat {\bf M}^{kj}_d\nonumber\\ 
+\left(4\lambda_{ad,cb}^{^{(45)}}-7\lambda_{ab,cd}^{^{(45)}}
\right)\widehat{\bf N}_a^{i\bf{T}}\widehat {\bf M}_{bi}\widehat{\bf
N}_{cjk}^{\bf{T}}\widehat 
{\bf M}^{jk}_d
+\left(4\lambda_{ad,cb}^{^{(45)}}+\lambda_{ab,cd}^{^{(45)}}
\right)\widehat{\bf N}_a^{\bf{T}}\widehat {\bf M}_b\widehat{\bf
N}_{cij}^{\bf{T}}\widehat {\bf M}^{ij}_d\nonumber\\
+\frac{1}{4}\lambda_{ab,cd}^{^{(45)}} 
[-8\epsilon^{ijklm}\widehat{\bf N}_{aij}^{\bf{T}}\widehat
{\bf M}_{bk}\widehat{\bf
N}_{clm}^{\bf{T}}\widehat {\bf M}_d
-8\epsilon_{ijklm}\widehat{\bf N}_a^{\bf{T}}\widehat
{\bf M}^{ij}_b\widehat{\bf
N}^{k\bf{T}}_c\widehat {\bf M}^{lm}_d                                                          
-16\widehat{\bf N}_{aik}^{\bf{T}}\widehat {\bf M}^{kj}_b\widehat{\bf
N}_{cjl}^{\bf{T}}\widehat 
{\bf M}^{li}_d\nonumber\\
+3\widehat{\bf N}_{aij}^{\bf{T}}\widehat {\bf M}^{ij}_b\widehat{\bf
N}_{ckl}^{\bf{T}}\widehat {\bf M}^{kl}_d
+24\widehat{\bf N}_a^{\bf{T}}\widehat {\bf M}_b\widehat{\bf
N}^{i\bf{T}}_c\widehat {\bf M}_{di}-20\widehat{\bf N}_a^{\bf{T}}
\widehat {\bf M}_b\widehat{\bf N}^{\bf{T}}_c\widehat {\bf
M}_d]\nonumber\\
\end{eqnarray}
\begin{eqnarray}
{\cal I}_{210}=-\frac{1}{24}[4\left(-18\lambda_{ad,cb}^{^{(210)}}-25
\lambda_{ab,cd}^{^{(210)}}
\right)\widehat{\bf N}^{i\bf{T}}_a\widehat {\bf M}_{bi}\widehat{\bf
N}^{j\bf{T}}_c\widehat {\bf M}_{dj}
+16\left(\lambda_{ad,cb}^{^{(210)}}+5\lambda_{ab,cd}^{^{(210)}}
\right)\widehat{\bf N}^{i\bf{T}}_a\widehat {\bf M}_{bj}\widehat{\bf
N}_{cik}^{\bf{T}}\widehat {\bf M}^{kj}_d\nonumber\\
+12\left(-2\lambda_{ad,cb}^{^{(210)}}+3\lambda_{ab,cd}^{^{(210)}}
\right)\widehat{\bf N}_a^{i\bf{T}}\widehat {\bf M}_{bi}\widehat{\bf
N}_{cjk}^{\bf{T}}\widehat
{\bf M}^{jk}_d
+4\left(-6\lambda_{ad,cb}^{^{(210)}}+\lambda_{ab,cd}^{^{(210)}}
\right)\widehat{\bf N}_a^{\bf{T}}\widehat {\bf M}_b\widehat{\bf
N}_{cij}^{\bf{T}}\widehat {\bf M}^{ij}_d\nonumber\\
+\left(8\lambda_{ad,cb}^{^{(210)}}+25\lambda_{ab,cd}^{^{(210)}}\right)
\widehat{\bf N}_{aij}^{\bf{T}}\widehat {\bf M}^{ij}_b\widehat{\bf
N}_{ckl}^{\bf{T}}\widehat {\bf M}^{kl}_d
+8\left(8\lambda_{ad,cb}^{^{(210)}}+\lambda_{ab,cd}^{^{(210)}}
\right)                                                         
\widehat{\bf N}_a^{\bf{T}}\widehat {\bf M}_b\widehat{\bf
N}^{i\bf{T}}_c\widehat {\bf M}_{di}]\nonumber\\                                      
+4\lambda_{ab,cd}^{^{(210)}}
\{-\epsilon^{ijklm}\widehat{\bf N}_{aij}^{\bf{T}}\widehat
{\bf M}_{bk}\widehat{\bf
N}_{clm}^{\bf{T}}\widehat {\bf M}_d
-\epsilon_{ijklm}\widehat{\bf N}_a^{\bf{T}}\widehat
{\bf M}^{ij}_b\widehat{\bf
N}^{k\bf{T}}_c\widehat {\bf M}^{lm}_d
-2\widehat{\bf N}_{aik}^{\bf{T}}\widehat {\bf M}^{kj}_b\widehat{\bf
N}_{cjl}^{\bf{T}}\widehat
{\bf M}^{li}_d\nonumber\\ 
+5\widehat{\bf N}_a^{\bf{T}}
\widehat {\bf M}_b\widehat{\bf N}^{\bf{T}}_c\widehat {\bf
M}_d\}]\nonumber\\
\end{eqnarray}
where
\begin{eqnarray}
\lambda_{ab,cd}^{^{(1)}}=
h_{ab}^{^{(1)}}h_{cd}^{^{(1)}}f_{_{{\cal X}}}^{^{(1)}}\left[\left({\cal
M}^{^{(1)}}+{\cal M}^{^{(1)\bf{T}}}\right)^{-1}\left\{{\cal
M}^{^{(1)}}
\left({\cal M}^{^{(1)}}+{\cal
M}^{^{(1)\bf{T}}}\right)^{-1}-\bf{1}\right\}
\right]_{{\cal X}{\cal X}'}f_{_{{\cal X}'}}^{^{(1)}}\nonumber\\            
\lambda_{ab,cd}^{^{(45)}}=
h_{ab}^{^{(45)}}h_{cd}^{^{(45)}}f_{_{{\cal Y}}}^{^{(45)}}\left[\left({\cal
M}^{^{(45)}}+{\cal M}^{^{(45)\bf{T}}}\right)^{-1}\left\{{\cal
M}^{^{(45)}}
\left({\cal M}^{^{(45)}}+{\cal
M}^{^{(45)\bf{T}}}\right)^{-1}-\bf{1}\right\}
\right]_{{\cal Y}{\cal Y}'}f_{_{{\cal Y}'}}^{^{(45)}}\nonumber\\
\lambda_{ab,cd}^{^{(210)}}=
h_{ab}^{^{(210)}}h_{cd}^{^{(210)}}f_{_{{\cal Z}}}^{^{(210)}}\left[\left({\cal
M}^{^{(210)}}+{\cal M}^{^{(210)\bf{T}}}\right)^{-1}\left\{{\cal
M}^{^{(210)}}
\left({\cal M}^{^{(210)}}+{\cal
M}^{^{(210)\bf{T}}}\right)^{-1}-\bf{1}\right\}
\right]_{{\cal Z}{\cal Z}'}f_{_{{\cal Z}'}}^{^{(210)}}.\nonumber\\
\end{eqnarray}     
 The exact same technique can be used to compute the quartic couplings
 of the form $[1616]_{10}[1616]_{10}$, $[1616]_{120}[1616]_{120}$,
 and  $[1616]_{\overline{126}}[1616]_{126}$ 
  arising from the elimination of the 10 plet, the 120 plet 
  and the $\overline{126}$ plet  
 of heavy Higgs using the cubic couplings already derived in Ref.\cite{ns}.
 Similarly one can compute   
 $[\overline{16}\overline{16}][\overline{16}\overline{16}]$
 and $[{16}16][\overline{16}\overline{16}]$ couplings using
 the technique above.                                            
As mentioned in Sec.1 the chirality $(+)$ fields, $16_{(+)a}$ can be
identified with the three generations of quarks and leptons and right handed
neutrino fields, and the chirality $(-)$ field, $\overline {16}_-$ can be
identified as a Higgs multiplet, $\overline {16}_H$.  With the above
identification one finds that there exist dimension five operators of the
type  $[16_a\overline {16}_H]_1[16_b\overline {16}_H]_1$,
$[16_a\overline {16}_H]_{45}[16_b\overline {16}_H]_{45}$, and
$[16_a\overline {16}_H]_{210}[16_b\overline {16}_H]_{210}$, which can be
computed by using the analysis given above. Similarly, one has operators
of the type $[16_a16_b]_{10}[\overline {16}_H\overline {16}_H]_{10}$,
$[16_a16_b]_{120}[\overline {16}_H\overline {16}_H]_{120}$, 
$[16_a16_b]_{\overline {126}}[\overline {16}_H\overline {16}_H]_{126}$ which
can be computed by similar techniques using the results of Ref.[5]. After
spontaneous breaking, the Higgs multiplet can develop vacuum expectation
values generating mass terms for some of the quark, lepton and neutrino
fields. An example of such operators can be found in Ref.\cite{bpw}. 
A further discussion of model building is given in Sec.7.
\section{Vector Couplings}
For the construction of couplings of vector fields with 
 $16_{\pm}$ plets it is natural to consider the couplings of the 
 1 and 45 vector fields as abelian and Yang-Mills gauge interactions.
 However, one cannot do the same for the $\overline{16}_{\pm}16_{\pm}210$
 couplings. These couplings cannot be treated as gauge couplings
 as there are no corresponding Yang-Mills interactions for the 210 plet.
 For this reason we focus here first on the computation of the gauge
 couplings of the 1 and 45 plet of vector fields. The supersymmetric
 kinetic energy and gauge couplings of the chiral superfield 
 $\hat{\phi}$ can be written in the usual superfield notation 
\begin{equation}
 \int d^4\theta~tr(\hat{\phi}^{\dagger}e^{g\hat{V}}\hat{\phi})
\end{equation}
where $\hat{V}$ is the Lie valued vector superfield. Similarly the 
supersymmetric Yang-Mills part of the Lagrangian can be gotten
from 

\begin{equation}
 \int d^2\theta~tr(W^{\alpha}W_{\alpha}))
+ \int d^2\bar{\theta}~
tr(\overline{W}_{\dot{\alpha}}\overline{W}^{\dot{\alpha}})  
\end{equation}
 where  $W_{\alpha}$ is the field strength
 chiral spinor superfield. Since supersymmetry does not play any special
 role in the analysis of SO(10) Clebsch-Gordon co-efficients,
 we will display in the analysis here only the parts of the Lagrangian
 relevant for our discussion. Thus the interactions of the 
 $16_+$ with gauge vectors for the 1 and 45 plet cases are
 given by

\begin{equation}
{\mathsf
L}^{(1)}_{++}=g^{^{(1)}}_{ab}<\Psi_{(+)a}|\gamma^0\gamma^A|\Psi_{(+)b}>\Phi_A
\end{equation}
\begin{equation}
{\mathsf L}^{(45)}_{++}=\frac{1}{i}\frac{1}{2!}g^{^{(45)}}_{ab}<\Psi_{(+)a}|\gamma^0\gamma^A
\Sigma_{\mu\nu}|\Psi_{(+)b}>\Phi_{A\mu\nu}
\end{equation}
where $\gamma^A (A,B=0-3)$ spans the Clifford algebra
associated with the Lorentz group, $g$'s are the gauge coupling constants,
and $\Phi_A$ and $\Phi_{A\mu\nu}$ are gauge tensors of dimensionality
1 and 45, respectively.
Similarly one defines ${\mathsf L}^{(1)}_{--}$,
${\mathsf L}^{(45)}_{--}$ with 
 $\Psi_{+}$ replaced by $\Psi_{-}$ in Eqs.(39) and (40).
 
 We first present the result of the trivial 
${\overline {16}\times 16\times 1}$ couplings. Eqs.(39) 
and (13) at once give
\begin{equation}
{\mathsf L}^{(1)}_{++}=g^{^{(1)}}_{ab}\left(\overline
{\bf M}_a\gamma^A{\bf M}_b+\frac{1}{2}\overline
{\bf M}_{aij}\gamma^A{\bf M}_b^{ij}+\overline {\bf M}_a^i\gamma^A{\bf M}_{bi}
\right){\mathsf G}_A.
\end{equation}
The barred matter fields are
defined so that $~\overline {\bf M}_{ij}={\bf M}_{ij}^{\dagger}\gamma ^0$
etc.  

A  similar analysis gives ${\mathsf   L}^{(1)}_{--}$ and one has                
\begin{eqnarray}
{\mathsf                                                                    
L}^{(1)}_{--}=g^{^{(1)}}_{ab}<\Psi_{(-)a}|\gamma^0\gamma^A|\Psi_{(-)b}>\Phi_A
~~~~~~~~~~~~~~~~~~~~~~~~\nonumber\\
=g^{^{(1)}}_{ab}\left(\overline
{\bf N}_a\gamma^A{\bf N}_b+\frac{1}{2}\overline
{\bf N}^{ij}_a\gamma^A{\bf N}_{bij}+\overline {\bf N}_{ai}\gamma^A{\bf
N}_b^i
\right){\mathsf G}_A.  
\end{eqnarray} 
We next discuss the couplings of the 45 plet gauge tensor
$\Phi_{A\mu\nu}$ whose decomposition in terms of reducible SU(5)  
tensors can be written similar to Eq.(16). This can be further 
reduced into irreducible parts similar to Eq.(17) by  
\begin{eqnarray}
\Phi_{Ac_n\overline c_n}={\mathsf g}_A;~~~\Phi_{Ac_i\overline c_j}={\mathsf g}_{Aj}^i+
\frac{1}{5}\delta_j^i{\mathsf g}_A;~~~
\Phi_{Ac_ic_j}={\mathsf g}_A^{ij};~~~\Phi_{A\overline c_i\overline c_j}={\mathsf g}_{Aij}
\end{eqnarray}
and normalized so that
\begin{eqnarray}
{\mathsf g}_A=2\sqrt 5 {\mathsf G}_A;~~~{\mathsf g}_{Aij}=\sqrt 2 {\mathsf G}
_{Aij};~~~
{\mathsf g}_A^{ij}=\sqrt 2 {\mathsf G}_A^{ij};~~~{\mathsf g}_{Aj}^i=\sqrt{2}
{\mathsf G}_{Aj}^i.
\end{eqnarray}
The kinetic energy 
for the 45-plet is given by $-\frac{1}{4}{\cal F}_{\mu\nu}^
{AB}{\cal F}_{AB\mu\nu}$, 
 where ${\cal F}_{\mu\nu}^{AB}$ is the 45 of SO(10) field strength tensor.
In terms of the redefined fields, 45-plet's kinetic energy takes the form
\begin{equation}
{\mathsf L}_{kin}^{45-gauge}=-\frac{1}{2}{\cal G}_{AB}{\cal
G}^{AB\dagger}-\frac{1}{2!}
\frac{1}{2}{\cal G}^{ABij}{\cal G}_{AB}^{ij\dagger}
-\frac{1}{2!}\frac{1}{2}{\cal G}_j^{ABi}{\cal G}_{ABi}^{j}
\end{equation}
where ${\cal F}_{\mu\nu}^{AB}$ is the 45 of SO(10) field strength tensor.
As mentioned in the beginning of this section we do not exhibit the 
gaugino and D terms needed
for supersymmetry  since their normalization is fixed 
relative to terms exhibited in Eq.(45).  
Using  Eqs.(40), (16) and the above normalizations we find 
\begin{eqnarray}
{\mathsf L}^{(45)}_{++}=g^{^{(45)}}_{ab}[\sqrt 5\left(-\frac{3}{5}\overline
{\bf M}_a^i\gamma^A{\bf M}_{bi}+\frac{1}{10}\overline {\bf M}_{aij}
\gamma^A{\bf M}_b^{ij}+
\overline{\bf M}_a\gamma^A{\bf M}_b\right){\mathsf G}_{A}\nonumber\\
+{\frac{1} {\sqrt 2}}\left(\overline {\bf M}_a \gamma^{A} {\bf M}_b^{lm}
+{\frac{1}{2}}\epsilon^{ijklm}\overline {\bf M}_{aij}\gamma^A{\bf M}_{bk}\right) 
{\mathsf G}_{Alm}\nonumber\\
-\frac{1}{\sqrt 2}\left(\overline
{\bf M}_{alm}\gamma^A{\bf M}_b+\frac{1}{2}\epsilon_{ijklm}
\overline {\bf M}_a^i\gamma^A{\bf M}_b^{jk}\right){\mathsf
G}_A^{lm}\nonumber\\
+\sqrt{2}\left(\overline {\bf M}_{aik}\gamma^A{\bf M}_b^{kj}+\overline
{\bf M}_a^j\gamma^A{\bf M}_{bi}\right){\mathsf G}_{Aj}^i]. 
\end{eqnarray}
A similar analysis gives
\begin{eqnarray}
{\mathsf
L}^{(45)}_{--}=\frac{1}{i}
 \frac{1}{2!}g^{^{(45)}}_{ab}<\Psi_{(-)a}|\gamma^0\gamma^A
\Sigma_{\mu\nu}|\Psi_{(-)b}>\Phi_{A\mu\nu}~~~~~~~~~~~~~~~~~~~~~~~~~~\nonumber\\
=g^{^{(45)}}_{ab}[\sqrt 5\left(\frac{3}{5}\overline
{\bf N}_{ai}\gamma^A{\bf N}_b^i-\frac{1}{10}\overline {\bf N}_a^{ij}
\gamma^A{\bf N}_{bij}-
\overline{\bf N}_a\gamma^A{\bf N}_b\right){\mathsf G}_{A}\nonumber\\
+{\frac{1} {\sqrt 2}}\left(\overline {\bf N}_a^{lm} \gamma^{A} {\bf N}_b
+{\frac{1}{2}}\epsilon^{ijklm}\overline {\bf N}_{ai}\gamma^A{\bf
N}_{bjk}\right)
{\mathsf G}_{Alm}\nonumber\\
-\frac{1}{\sqrt 2}\left(\overline
{\bf N}_a\gamma^A{\bf N}_{blm}+\frac{1}{2}\epsilon_{ijklm}
\overline {\bf N}_a^{ij}\gamma^A{\bf N}_b^k\right){\mathsf
G}_A^{lm}\nonumber\\
-\sqrt{2}\left(\overline {\bf N}_a^{jk}\gamma^A{\bf N}_{bki}+\overline
{\bf N}_{ai}\gamma^A{\bf N}_b^j\right){\mathsf G}_{Aj}^i].       
\end{eqnarray}

We discuss now the 210 vector multiplet. This vector mutiplet 
is not a gauge multiplet with the usual Yang-Mills interactions.
This makes the multiplet rather pathological 
and it cannot be treated in a normal fashion. 
Specifically Eq.(37) is not valid for this case 
in any direct fashion.
However, for the 
sake of completeness, we present here the SO(10) globally 
invariant couplings corresponding  to Eq.(40). Thus we
 have 
\begin{equation}
{\mathsf L}^{(210)}_{++}=\frac{1}{4!}g^{^{(210)}}_{ab}<\Psi_{(+)a}
|\gamma^0\gamma^A\Gamma_{[\mu}\Gamma_{\nu}\Gamma_{\rho}
\Gamma_{\lambda]}|\Psi_{(+)b}>\Phi_{A\mu\nu\rho\lambda}.
\end{equation}  
To compute the couplings we carry out  expansions similar to
Eqs.(22) and (23) and to normalize the fields 
we carry out a field redefinition 
%\vfill \eject
\begin{eqnarray}
{\mathsf g}_A=4\sqrt{\frac{10}{3}}{\mathsf G}_A;~~~{\mathsf g}_A^i=
8\sqrt{6}{\mathsf G}_A^i;
~~~{\mathsf g}_{Ai}=8\sqrt{6}{\mathsf G}_{Ai}\nonumber\\
{\mathsf g}_A^{ij}=\sqrt{2}{\mathsf G}_A^{ij};~~~{\mathsf g}_{Aij}=
\sqrt{2}{\mathsf G}_{Aij};
~~~{\mathsf g}_{Aj}^i
=\sqrt{2}{\mathsf G}_{Aj}^i\nonumber\\
{\mathsf g}_{Al}^{ijk}=\sqrt{\frac{2}{3}}{\mathsf G}_{Al}^{ijk};
~~~{\mathsf g}_{Ajkl}^i
=\sqrt{\frac{2}{3}}{\mathsf G}_{Ajkl}^i;~~~{\mathsf g}_{Akl}^{ij}=\frac{2}
{\sqrt{3}}{\mathsf G}_{Akl}^{ij}
\end{eqnarray}
so that the 210-plet's kinetic energy
$-\frac{1}{4}{\cal F}_{\mu\nu\rho\lambda}^{AB}{\cal F}_{AB\mu\nu\rho\lambda}$
takes the form
\begin{eqnarray}
{\mathsf L}_{kin}^{210-gauge}=-\frac{1}{2}{\cal G}_{AB}{\cal
G}^{AB\dagger}-\frac{1}{2}
{\cal G}_{AB}^i{\cal G}^{ABi\dagger}
-\frac{1}{2!}\frac{1}{2}{\cal G}_{AB}^{ij}{\cal G}^{ABij\dagger}\nonumber\\        
-\frac{1}{2!}\frac{1}{2}{\cal G}_{ABj}^i{\cal G}_i^{ABj}
-\frac{1}{3!}\frac{1}{2}{\cal G}_{ABl}^{ijk}
{\cal G}_l^{ABijk\dagger}
-\frac{1}{2!}\frac{1}{2!}\frac{1}{2}{\cal G}_{ABkl}^{ij}{\cal
G}_{kl}^{ABij\dagger}
.\end{eqnarray}

\noindent
As discussed above, the 210 vector multiplet is not a gauge
multiplet and thus the quantity ${\cal G}_{AB}$ is just
an ordinary curl.  
Using Eqs.(48), (22) and the normalizations of Eq.(49) one can compute 
${\mathsf L}^{(210)}_{++}$. One finds
\begin{eqnarray}
{\mathsf
L}^{(210)}_{++}=\frac{1}{\sqrt{6}}g^{^{(210)}}_{ab}
[\sqrt{5}\left
(\overline {\bf M}_a\gamma^A{\bf M}_b-{\frac{1}{10}}\overline {\bf M}_{aij}
\gamma^A{\bf M}_b^{ij}
+\frac{1}{5}\overline {\bf M}_a^i\gamma^A{\bf M}_{bi}\right){\mathsf G}_A\nonumber\\
+\frac{\sqrt{3}}{2}\left(-\overline{\bf M}_a\gamma^A{\bf M}_b^{lm}
+\frac{1}{6}\epsilon^{ijklm}\overline{\bf M}_{aij}
\gamma^A{\bf M}_{bk}\right){\mathsf G}_{Alm}\nonumber\\
+\frac{\sqrt{3}}{2}\left(-\overline{\bf M}_{alm}\gamma^A{\bf M}_b
+\frac{1}{6}\epsilon_{ijklm}\overline{\bf M}_a^i
\gamma^A{\bf M}_b^{jk}\right){\mathsf G}_A^{lm}\nonumber\\
+\sqrt{3}\left(-\overline {\bf M}_a^j\gamma^A{\bf M}_{bi}
+\frac{1}{3}\overline{\bf M}_{aik}\gamma^A{\bf M}_b^{kj}\right){\mathsf G}_{Aj}^i
\nonumber\\
-\frac{1}{3}\epsilon^{ijklm}\overline {\bf M}_{ain}
\gamma^A{\bf M}_{bj}{\mathsf G}_{Aklm}^n
+\frac{1}{3}\epsilon_{ijklm}\overline {\bf M}_a^i
\gamma^A{\bf M}_b^{jn}{\mathsf G}_{An}^{klm}\nonumber\\
-\frac{1}{\sqrt{2}}\overline {\bf M}_{aij}
\gamma^A{\bf M}_b^{kl}{\mathsf G}_{Akl}^{ij}
+2\overline {\bf M}_a^i\gamma^A{\bf M}_bG_{Ai}
+2\overline {\bf M}_a\gamma^A{\bf M}_{bi}{\mathsf G}_A^i].
\end{eqnarray}
A similar analysis gives 
\begin{eqnarray}
{\mathsf L}^{(210)}_{--}=\frac{1}{4!}g^{^{(210)}}_{ab}<\Psi_{(-)a}
|\gamma^0\gamma^A\Gamma_{[\mu}\Gamma_{\nu}\Gamma_{\rho}
\Gamma_{\lambda]}|\Psi_{(-)b}>\Phi_{A\mu\nu\rho\lambda}~~~~~~~~~~~~~~~~~~~~
\nonumber\\
=\frac{1}{\sqrt{6}}g^{^{(210)}}_{ab}
[\sqrt{5}\left
(\overline {\bf N}_a\gamma^A{\bf N}_b-{\frac{1}{10}}\overline {\bf N}_a^{ij}
\gamma^A{\bf N}_{bij}
+\frac{1}{5}\overline {\bf N}_{ai}\gamma^A{\bf N}_b^i\right){\mathsf
G}_A\nonumber\\
+\frac{\sqrt{3}}{2}\left(\overline{\bf N}_a^{lm}\gamma^A{\bf N}_b
-\frac{1}{6}\epsilon^{ijklm}\overline{\bf N}_{ai}
\gamma^A{\bf N}_{bjk}\right){\mathsf G}_{Alm}\nonumber\\
+\frac{\sqrt{3}}{2}\left(\overline{\bf N}_a\gamma^A{\bf N}_{blm}
-\frac{1}{6}\epsilon_{ijklm}\overline{\bf N}_a^{ij}
\gamma^A{\bf N}_b^{k}\right){\mathsf G}_A^{lm}\nonumber\\
+\sqrt{3}\left(-\overline {\bf N}_{ai}\gamma^A{\bf N}_b^j
+\frac{1}{3}\overline{\bf N}_a^{jk}\gamma^A{\bf N}_{bki}\right){\mathsf
G}_{Aj}^i\nonumber\\
+\frac{1}{6}\epsilon^{ijklm}\overline {\bf N}_{an}
\gamma^A{\bf N}_{bij}{\mathsf G}_{Aklm}^n
+\frac{1}{6}\epsilon_{ijklm}\overline {\bf N}_a^{ij}
\gamma^A{\bf N}_b^{n}{\mathsf G}_{An}^{klm}\nonumber\\
-\frac{1}{\sqrt{2}}\overline {\bf N}_a^{kl}
\gamma^A{\bf N}_{bij}{\mathsf G}_{Akl}^{ij}
+2\overline {\bf N}_a\gamma^A{\bf N}_b^iG_{Ai}
+2\overline {\bf N}_{ai}\gamma^A{\bf N}_b{\mathsf G}_A^i].
\end{eqnarray}  
Supersymmetrizations of Eqs.(51) and (52) requires that we deal with a
massive vector multiplet and this topic will be dealt elsewhere\cite{pr}.

%%%%%%%%%%%%%%%%%%%%%%%%%%%%%%%%%%%%%%%%%%%%%%%%%%%%%%%%%%%newsection
 \section{Possible role of large tensor 
    representations in model building}
 Most of the model building in SO(10) has occured using small 
 Higgs representations\cite{pati} and large
 representations are generally avoided as they lead to 
non-perturbative physics above the grand unified scale. 
However, for the purposes of physics below the grand unified 
scale, the existence of non-perturbativity above the unified scale
is not a central concern since the region above this scale in any case
cannot
be fully understood without taking into account quantum gravity effects.
Thus there is no fundamental reason not to consider 
 model building which allows for couplings with
large tensor representations. Indeed large tensor representations
have some very interesting and desirable features. Thus, for example,
if the $\overline{126}$ develops a VEV in the direction of 
$\overline{45}$ of $SU(5)$
one can get  the ratio 3:1 in the "22" element of the lepton vs. 
the down quark sector in a natural fashion  as desired in the Georgi-Jarlskog textures\cite{gj}.
A similar 3:1 ratio also appears in the 120 plet couplings. 
Because of this feature the tensor representations $120$ and $\overline{126}$ 
have already  
appeared in several analyses of lepton and quark textures\cite{gn}. 
Further, it was pointed out in Ref.\cite{ns} that the tensor 
representation $\overline{126}$ may also play a  role in suppressing
proton decay arising from dimension five operators in supersymmetric 
models. 
This is so because couplings involving $\overline{126}$ plet of
Higgs to 16 plet of matter 
do not give rise to dimension five operators. The result derived
here including the computation of cubic and quartic couplings may find
application also in the study of neutrino masses and mixings.
Thus, for example, one may consider contributions to the neutrino
mass (N) and to the up quark mass (U) from the contraction 
$[16_a\overline{16}_H]_{45}[16_b\overline{16}_H]_{45}$.
 From Eq.(34) we 
find that a contribution to N arises from the fifth term in the
last parentheses of Eq.(34) while the contribution to U arises from
the second term in the last parentheses of Eq.(34). Now comparing the 
above with Eq.(8) of Ref.\cite{ns} for the 10 plet Higgs coupling
which gives a N:U ratio of 1:1 we find that the two couplings
refered to above in Eq.(34) give N:U=3:8 in agreement with
Ref.\cite{babu}.  
Regarding the 210 dimensional tensor, such a mutiplet could play a role
in the quark-lepton and neutrino mass textures. The role of a
210 dimensional vector multiplet is less clear.  One possible way
it may surface in low energy physics is as a condensate field. However,
this topic needs further exploration. 
A more detailed discussion of model building including large tensor 
representations is under investigation.

\section{Conclusion}
In this paper we  have given a complete determination of the $SO(10)$ 
invariant couplings $\overline {16}_{\mp}-16_{\pm}-1$,
$\overline {16}_{\mp}-16_{\pm}-45$ and $\overline {16}_{\mp}-16_{\pm}-210$
 in the superpotential in their $SU(5)$ decomposed form.
 Further, we have computed all the allowed quartic interactions in
 the superpotential of the type 
 $\overline{16}_{\mp}16_{\pm}\overline{16}_{\mp}16_{\pm}$. We also 
 exhibited a technique which is much simpler and involves elimination
 of heavy fields in cubic couplings in their SU(5) decomposed form.
  These techniques can be directly applied to the 
 computation of quartic couplings of the type $16_+16_+16_+16_+$ using 
 cubic couplings involving $16_+16_+$ with the 10, 120 and $\overline{126}$
 tensor multiplets which have already been computed in the work of Ref.\cite{ns}.
   An analysis of vector couplings
 involving the SO(10) vector mutiplets 1, 45 and 210 was also given. 
  In all of our analysis we have made explicit use of
the theorem developed in Ref\cite{ns} on the decomposition of 
SO(10) vertices which allows  the complete determination
of the couplings with large tensor representations.
 It would be very 
straightforward now to expand all the $SU(5)$ invariants in 
terms of $SU(3)_C\times 
SU(2)_L\times U(1)_Y$ invariants using the particle assignments given by
Eq.(7) and an example of this is given in Appendix B. We also discussed
in this paper some interesting features of large tensor representations
 and the role they may play in future model building.

\section{Acknowledgements}
We have enjoyed discussions with Luis Alvarez-Gaume, Kaladi Babu,
Stuart Raby and Gabriele Veneziano. 
One of us (PN) wishes to thank
the Physics Institute at the University of Bonn
and the Max Planck Institute, Heidelberg, where a part of this work 
was carried out, for hospitality
and acknowledges support from the Alexander von Humboldt Foundation.
R.S. was supported by a GAANN Fellowship from the U.S. Department
of Education and ASCC at Northeastern University.
This research was supported in part by NSF grant PHY-9901057.\\
\section{Appendix A}
We expand here on the technique for the elimination of heavy fields
for the case when the fields belong to a large tensor representation.
There are infact three approaches one can use in affecting this 
elimination. The first one is the direct approach where one eliminates
the heavy large Higgs representation in its SO(10) form. 
While this is the most straightforward  approach the disadvantage is
that the analysis of dimension 4 operators cannot be directly 
made use of and one has to carry out the entire computation from
scratch. 
An alternative possibility is that one utilizes the result of
computations of dimension 4 operators already done to compute 
dimension five operators. In this case, however, since all the 
heavy Higgs fields are in their SU(5) irreducible representations
the elimination of such fields would involve cross cancellations 
which are quite delicate. Thus, for example, in its SU(5) decomposition
$210=1+5+\bar 5+10+ \overline{10}+24+40+\overline{40}+75$ and 
elimination of these involve cancellations between the 10 and the
40 plet contributions, between the $\overline {10}$ and the
$\overline {40}$ plet contributions, and between the 1, 24 and 75 plet
contributions. Such cancellations make the analysis tedious once
again. It turns out that there is yet a third possibility which 
is to derive the dimension 4 operators in SU(5) decomposition 
leaving the SU(5) fields in their reducible form where possible, 
i.e., to use Eq.(13) without further reduction of the tensor fields in
their irreducible components. Thus, for example, in this case one
would carry out the following SU(5) decomposition of the SO(10)
tensor, $210=5+\bar 5+50+\overline{50}+100$  where 
$ 50, \overline{50},100$ are reducible SU(5) representations.
 After computing the dimension 4 operators in terms of these tensors 
  one eliminates them. This procedure has the advantage of having 
  the cancellations of procedure 2 already built in. 
We give now more details of the three approaches. 

We begin by discussion of the first approach where one eliminates
the heavy fields in the superpotential before one carries out
an SU(5) decomposition. Here on using the flatness conditions one
finds

 \begin{equation}  
{\cal I}_{1}=2\lambda_{ab,cd}^{^{(1)}}
<\widehat{\Psi}^*_{(-)a}|B|\widehat{\Psi}_{(+)b}><\widehat{\Psi}^*_{(-)c}|B|
\widehat{\Psi}_{(+)d}> 
\end{equation} 
 \begin{eqnarray}
{\cal I}_{45}=-\frac{1}{2}\lambda_{ab,cd}^{^{(45)}}
[<\widehat{\Psi}^*_{(-)a}|B\Gamma_{\mu}\Gamma_{\nu}|\widehat{\Psi}_{(+)b}>
<\widehat{\Psi}^*_{(-)c}|B\Gamma_{\mu}\Gamma_{\nu}|\widehat{\Psi}_{(+)d}>\nonumber\\
-10<\widehat{\Psi}^*_{(-)a}|B|\widehat{\Psi}_{(+)b}><\widehat{\Psi}^*_{(-)c}|B|
\widehat{\Psi}_{(+)d}>]
\end{eqnarray}
Expansion in oscillator modes gives 
\begin{eqnarray}
{\cal I}_{45}
=\lambda_{ab,cd}^{^{(45)}}              
[-4<\widehat{\Psi}^*_{(-)a}|Bb_ib_j|\widehat{\Psi}_{(+)b}>
<\widehat{\Psi}^*_{(-)c}|Bb_i^{\dagger}b_j^{\dagger}|\widehat{\Psi}_{(+)d}>
~~~\nonumber\\
+4<\widehat{\Psi}^*_{(-)a}|Bb_i^{\dagger}b_j|\widehat{\Psi}_{(+)b}>
<\widehat{\Psi}^*_{(-)c}|Bb_j^{\dagger}b_i|\widehat{\Psi}_{(+)d}>
~~~\nonumber\\  
-4<\widehat{\Psi}^*_{(-)a}|Bb_n^{\dagger}b_n|\widehat{\Psi}_{(+)b}>
<\widehat{\Psi}^*_{(-)c}|B|\widehat{\Psi}_{(+)d}>~~~\nonumber\\
+5<\widehat{\Psi}^*_{(-)a}|B|\widehat{\Psi}_{(+)b}><\widehat{\Psi}^*_{(-)c}|B|
\widehat{\Psi}_{(+)d}>].~~~
 \end{eqnarray}  
 A  similar  analysis for the 210 plet field gives 
\begin {eqnarray}
{\cal I}_{210}=\frac{1}{288}\lambda_{ab,cd}^{^{(210)}}
[<\widehat{\Psi}^*_{(-)a}|B\Gamma_{\mu}\Gamma_{\nu}\Gamma_{\rho}
\Gamma_{\lambda}|\widehat{\Psi}_{(+)b}>
<\widehat{\Psi}^*_{(-)c}|B\Gamma_{\mu}\Gamma_{\nu}\Gamma_{\rho}
\Gamma_{\lambda}|            
\widehat{\Psi}_{(+)d}>\nonumber\\                                                                         
-52<\widehat{\Psi}^*_{(-)a}|B\Gamma_{\mu}\Gamma_{\nu}|\widehat{\Psi}_{(+)b}>
<\widehat{\Psi}^*_{(-)c}|B\Gamma_{\mu}\Gamma_{\nu}|\widehat{\Psi}_{(+)d}>
\nonumber\\                                                                         
+240<\widehat{\Psi}^*_{(-)a}|B|\widehat{\Psi}_{(+)b}><\widehat{\Psi}^*_{(-)c}
|B|\widehat{\Psi}_{(+)d}>]\nonumber\\
=-\frac{1}{18}\lambda_{ab,cd}^{^{(210)}}
[8<\widehat{\Psi}^*_{(-)a}|Bb_i^{\dagger}b_jb_kb_l|\widehat{\Psi}_{(+)b}>
<\widehat{\Psi}^*_{(-)c}|Bb_j^{\dagger}b_k^{\dagger}b_l^{\dagger}b_i
|\widehat{\Psi}_{(+)d}>
\nonumber\\
-6<\widehat{\Psi}^*_{(-)a}|Bb_i^{\dagger}b_j^{\dagger}b_kb_l
|\widehat{\Psi}_{(+)b}>                                       
<\widehat{\Psi}^*_{(-)c}|Bb_k^{\dagger}b_l^{\dagger}b_ib_j
|\widehat{\Psi}_{(+)d}>
\nonumber\\              
-2<\widehat{\Psi}^*_{(-)a}|Bb_ib_jb_kb_l
|\widehat{\Psi}_{(+)b}>
<\widehat{\Psi}^*_{(-)c}|Bb_i^{\dagger}b_j^{\dagger}b_k^{\dagger}
b_l^{\dagger}              
|\widehat{\Psi}_{(+)d}>
\nonumber\\
+24<\widehat{\Psi}^*_{(-)a}|Bb_i^{\dagger}b_j
|\widehat{\Psi}_{(+)b}>                   
<\widehat{\Psi}^*_{(-)c}|Bb_j^{\dagger}b_n^{\dagger}b_nb_i
|\widehat{\Psi}_{(+)d}> 
\nonumber\\                              
-12<\widehat{\Psi}^*_{(-)a}|Bb_i^{\dagger}b_j^{\dagger}
|\widehat{\Psi}_{(+)b}>
<\widehat{\Psi}^*_{(-)c}|Bb_n^{\dagger}b_nb_ib_j
|\widehat{\Psi}_{(+)d}>
\nonumber\\ 
-12<\widehat{\Psi}^*_{(-)a}|Bb_ib_j
|\widehat{\Psi}_{(+)b}>
<\widehat{\Psi}^*_{(-)c}|Bb_i^{\dagger}b_j^{\dagger}b_n^{\dagger}b_n
|\widehat{\Psi}_{(+)d}>
\nonumber\\  
-6<\widehat{\Psi}^*_{(-)a}|Bb_m^{\dagger}b_m
|\widehat{\Psi}_{(+)b}>
<\widehat{\Psi}^*_{(-)c}|Bb_n^{\dagger}b_n
|\widehat{\Psi}_{(+)d}>
\nonumber\\ 
-6<\widehat{\Psi}^*_{(-)a}|B
|\widehat{\Psi}_{(+)b}>
<\widehat{\Psi}^*_{(-)c}|Bb_m^{\dagger}b_n^{\dagger}b_nb_m
|\widehat{\Psi}_{(+)d}>
\nonumber\\
+18<\widehat{\Psi}^*_{(-)a}|Bb_ib_j
|\widehat{\Psi}_{(+)b}>
<\widehat{\Psi}^*_{(-)c}|Bb_i^{\dagger}b_j^{\dagger}
|\widehat{\Psi}_{(+)d}>
\nonumber\\
-18<\widehat{\Psi}^*_{(-)a}|Bb_i^{\dagger}b_j
|\widehat{\Psi}_{(+)b}>
<\widehat{\Psi}^*_{(-)c}|Bb_j^{\dagger}b_i
|\widehat{\Psi}_{(+)d}>
\nonumber\\
+24<\widehat{\Psi}^*_{(-)a}|B
|\widehat{\Psi}_{(+)b}>
<\widehat{\Psi}^*_{(-)c}|Bb_n^{\dagger}b_n
|\widehat{\Psi}_{(+)d}>
\nonumber\\ 
-15<\widehat{\Psi}^*_{(-)a}|B
|\widehat{\Psi}_{(+)b}>
<\widehat{\Psi}^*_{(-)c}|B
|\widehat{\Psi}_{(+)d}>].
\end {eqnarray}
 Although this is the most straigtforward technique, one has to
 carry out the entire analysis ab initio and can be very labor intensive
 for the case of large tensor representations.
 
We discuss now the second  approach where one decomposes the 
large tensor representations in its irreducible SU(5) components and
utilizes the results of the cubic superpotential already computed 
to derive  dimension five operators. For illustration we consider
the elimination of the 45 plet in the $\overline {16}-16-45$ coupling and
for simplicity we consider only one generation of Higgs.
 We begin by displaying the 45 plet mass term in terms of its irreducible 
SU(5) components 

\begin{eqnarray}
\frac{1}{2}{\cal M}^{^{(45)}}\Phi_{\mu\nu}\Phi_{\mu\nu}
=\frac{1}{2}{\cal M}^{^{(45)}}
\left[{\mathsf H}^{ij}{\mathsf H}_{ij}
-{\mathsf H}^i_j{\mathsf H}^j_i
-{\mathsf H}^2\right].
\end{eqnarray}
The superpotential is given by 
\begin{eqnarray}
{\mathsf W}^{(45)}_{-+}=
J^{(1/45)}{\mathsf H}
+J^{(\overline {10}/45)ij}{\mathsf H}_{ij}
+J^{(10/45)}_{ij}{\mathsf H}^{ij}
+J^{(24/45)j}_i{\mathsf H}_j^i 
\end{eqnarray}
where
\begin{eqnarray}
J^{(1/45)} =\sqrt{\frac{5}{2}}h_{ab}^{^{(45)}}\left(\frac{3}{5}\widehat
{\bf N}_a^{i\bf{T}}\widehat {\bf M}_{bi}+\frac{1}{10}\widehat {\bf
N}^{\bf{T}}_{aij}\widehat {\bf M}_b^{ij}-
\widehat {\bf N}_a^{\bf{T}}\widehat {\bf M}_b\right)\nonumber\\
J^{(\overline {10}/45)lm} =\frac{h_{ab}^{^{(45)}}}{\sqrt 2}\left(-\widehat
{\bf N}_a^{\bf{T}}\widehat {\bf M}_b^{lm}+\frac{1}{2}\epsilon^{ijklm}
\widehat
{\bf N}^{\bf{T}}_{alm}\widehat {\bf M}_{bk}\right)\nonumber\\
J^{(10/45)}_{lm} =\frac{h_{ab}^{^{(45)}}}{\sqrt2}\left(-\widehat
{\bf N}^{\bf{T}}_{alm}\widehat {\bf M}_b+
\frac{1}{2}\epsilon_{ijklm}\widehat                              
{\bf N}_a^{i\bf{T}}\widehat {\bf M}_b^{jk}\right)\nonumber\\
J^{(24/45)j}_i=\sqrt {2}h_{ab}^{^{(45)}}
\left(\widehat {\bf N}^{\bf{T}}_{aik}\widehat {\bf M}_b^{kj}-\widehat
{\bf
N}_a^{j\bf{T}}\widehat {\bf M}_{bi}\right).
\end{eqnarray}
Eliminating the irreducible SU(5) heavy Higgs fields through F-flatness
conditions taking care of the tracelessnes condition for $H_i^j$
one gets
\begin{eqnarray}
{\mathsf I}_{45}=\frac{1}{{10\cal M}^{^{(45)}}}
[5J^{(1/45)}J^{(1/45)}                   
-20J^{(\overline {10}/45)ij}J^{(10/45)}_{ij}\nonumber\\                    
+5J^{(24/45)j}_iJ^{(24/45)i}_j
-J^{(24/45)m}_mJ^{(24/45)n}_n].
\end{eqnarray} 
${\mathsf I}_{45}$ computed above is the same as ${\cal I}_{45}$ given by 
Eq.(34) using the direct method 
with $\frac{h_{ab}^{^{(45)}}h_{cd}^{^{(45)}}}{{\cal M}^{^{(45)}}}$
replaced by $-4\lambda_{ab,cd}^{^{(45)}}$. As pointed out in the
beginning of this appendix one has cancellations in this procedure
between the contributions arising from elimination of the 1 plet and
the 24 plet. Such cancellations become more abundant for the 210 
plet case.
Thus for this case it is more convenient to decompose the 210 plet into
reducible SU(5) tensors.  We begin by exhibiting the mass term
for this case

\begin{eqnarray}
\frac{1}{2}{\cal M}^{^{(210)}}\Phi_{\mu\nu\rho\lambda}\Phi_{\mu\nu\rho\lambda}
=\frac{1}{4}{\cal M}^{^{(210)}}\left[\frac{1}{4}{\mathsf K}^{ijkl}{\mathsf K}_{ijkl}
+{\mathsf K}^{jkl}_i{\mathsf K}_{jkl}^i
+\frac{3}{4}{\mathsf K}^{kl}_{ij}{\mathsf K}_{kl}^{ij}\right]
\end{eqnarray}
where ${\mathsf K}^{ijkl}$, ${\mathsf K}_{ijkl}$, ${\mathsf K}^{jkl}_i$,
${\mathsf K}_{jkl}^i$ and ${\mathsf K}_{kl}^{ij}$ are the $5$plet, 
$\bar 5$plet, $50$plet, $\overline {50}$plet and $100$plet
representations of SU(5). As before we keep only one generation of Higgs.
The superpotential ${\mathsf W}^{210)}_{-+}$ in this case may be 
written as

\begin{eqnarray}
{\mathsf W}^{210)}_{-+}=
J^{(\overline {5}/210)}_{ijkl}{\mathsf K}^{ijkl}
+J^{(5/210)ijkl}{\mathsf K}_{ijkl}
+J^{(50/210)l}_{ijk}{\mathsf K}_l^{ijk}
+J^{(\overline {50}/210)ijk}_l{\mathsf K}^l_{ijk}\nonumber\\
+J^{(50/210)}_{ij}{\mathsf K}^{ijn}_n
+J^{(\overline {50}/210)ij}{\mathsf K}^n_{ijn}
+J^{(100/210)ij}_{kl}{\mathsf K}^{kl}_{ij}
+J^{(100/210)j}_{i}{\mathsf K}^{in}_{jn}\nonumber\\
+J^{(100/210)}{\mathsf K}^{mn}_{mn}
\end{eqnarray}
where
\begin{eqnarray}
J_{ijkl}^{(\bar 5/210)}
=\frac{h_{ab}^{^{(210)}}}{24}
<\widehat{\Psi}^*_{(-)a}|Bb_i^{\dagger}b_j^{\dagger}b_k^{\dagger}
b_l^{\dagger}
|\widehat{\Psi}_{(+)b}>\nonumber\\
J^{(5/210)ijkl}=
\frac{h_{ab}^{^{(210)}}}{24}<\widehat{\Psi}^*_{(-)a}|Bb_ib_jb_kb_l   
|\widehat{\Psi}_{(+)b}>\nonumber\\
J^{(50/210)l}_{ijk}=\frac{h_{ab}^{^{(210)}}}{6}
<\widehat{\Psi}^*_{(-)a}|Bb_i^{\dagger}b_j^{\dagger}
b_k^{\dagger}b_l
|\widehat{\Psi}_{(+)b}>
\nonumber\\
J^{(\overline
{50}/210)jkl}_i=-\frac{h_{ab}^{^{(210)}}}{6}
<\widehat{\Psi}^*_{(-)a}|Bb_i^{\dagger}b_jb_kb_l|\widehat{\Psi}_{(+)b}>
\nonumber\\
J^{(50/210)}_{ij}=-\frac{h_{ab}^{^{(210)}}}{4}
<\widehat{\Psi}^*_{(-)a}|Bb_i^{\dagger}b_j^{\dagger}
|\widehat{\Psi}_{(+)b}>\nonumber\\
J^{(\overline {50}/210)ij}=\frac{h_{ab}^{^{(210)}}}{4}
<\widehat{\Psi}^*_{(-)a}|Bb_ib_j
|\widehat{\Psi}_{(+)b}>\nonumber\\
J^{(100/210)kl}_{ij}=\frac{h_{ab}^{^{(210)}}}{4}
\widehat{\Psi}^*_{(-)a}|Bb_i^{\dagger}b_j^{\dagger}b_kb_l
|\widehat{\Psi}_{(+)b}>\nonumber\\
J^{(100/210)j}_{i}= \frac{h_{ab}^{^{(210)}}}{2}
<\widehat{\Psi}^*_{(-)a}|Bb_i^{\dagger}b_j
|\widehat{\Psi}_{(+)b}>\nonumber\\
J^{(100/210)}=-\frac{h_{ab}^{^{(210)}}}{8}
<\widehat{\Psi}^*_{(-)a}|B
|\widehat{\Psi}_{(+)b}>.
\end{eqnarray}
Eliminating the reducible SU(5) Higgs fields through the F-flatness
condition we get
\begin{eqnarray}
{\mathsf I}_{210}=-\frac{1}{{3\cal M}^{^{(210)}}}[4
J^{(100/210)ij}_{kl}J^{(100/210)kl}_{ij}+8
J^{(100/210)mi}_{mj}J^{(100/210)j}_{i}\nonumber\\
+8J^{(100/210)mn}_{mn}J^{(100/210)}
+3J^{(100/210)j}_{i}J^{(100/210)i}_{j}\nonumber\\
+J^{(100/210)m}_{m}J^{(100/210)n}_{n}
+16J^{(100/210)m}_{m}J^{(100/210)} \nonumber\\
+40J^{(100/210)}J^{(100/210)}
+48J^{(5/210)ijkl}J^{({\overline 5}/210)}_{ijkl}\nonumber\\
+12J^{(\overline{50}/210)ijk}_lJ^{(50/210)l}_{ijk}
+12J^{(\overline{50}/210)mij}_mJ^{(50/210)}_{ij}\nonumber\\
+12J^{(\overline{50}/210)ij}J^{(50/210)m}_{ijm}
+12J^{(\overline{50}/210)ij}J^{(50/210)}_{ij}].
\end{eqnarray}
One may now check that ${\mathsf I}_{210}$ derived  above  coincides with 
${\cal I}_{210}$ given by
Eq.(56) using the direct method when we make the identification 
$\frac{h_{ab}^{^{(210)}}h_{cd}^{^{(210)}}}{{\cal
M}^{^{(210)}}}$ with $-4\lambda_{ab,cd}^{^{(210)}}$. 

\section{Appendix B}
In this appendix we expand some of the SO(10) interactions in
the familiar particle notation and exhibit the 
differences between some of the $\overline{16}-16-45$ and 
the $\overline{16}-16-210$
couplings. We start by looking at the gauge interactions of the 
24 plet of SU(5) 
in $\overline{16}-16-45$ coupling. We can read this off from the 
last term in Eq.(46). Disregarding the front factor, this term is of the
form
\begin{equation}
{\cal L}_{24/45}=g_{ab}^{^{(45)}}
\left(\overline {\bf M}_{aik}\gamma^A{\bf
M}_b^{kj}+\overline
{\bf M}_a^j\gamma^A{\bf M}_{bi}\right){\mathsf G}_{Aj}^i   
\end {equation}
An expansion of Eq.(65)  using the SM particle states defined 
by Eq.(7) gives 
\begin{eqnarray}
{\cal L}_{24/45}=g_{ab}^{^{(45)}} 
\sum_{x=1}^8\left[\overline U_a\gamma^A{\bf V}_A^{x}
\frac{\lambda_x}{2}U_b+\overline D_a\gamma^A{\bf V}_A^{x}\frac{\lambda_x}{2}
D_b\right]\nonumber\\ 
+g_{ab}^{^{(45)}}\sum_{y=1}^3\left[\left(\matrix{{\overline \nu}& 
{\overline E}^-}\right)_{aL}\gamma^A{\bf W}_A^{y}\frac{\tau_y}{2}
\left(\matrix{\nu\cr
E^-}\right)_{bL}
+\left(\matrix{{\overline U}&
{\overline D}}\right)_{aL}\gamma^A{\bf W}_A^{y}\frac{\tau_y}{2}
\left(\matrix{U\cr
D}\right)_{bL}\right]\nonumber\\
+g_{ab}^{^{(45)}} 
\sqrt{\frac{3}{5}}[-\frac{1}{2}\left({\overline E}_{aL}^-\gamma^A
{\bf B}_AE^-_{bL}+{\overline \nu}_{aL}\gamma^A
{\bf B}_A\nu_{bL}\right)
+\frac{1}{6}\left({\overline U}_{aL}^-\gamma^A
{\bf B}_AU_{bL}+{\overline D}_{aL}\gamma^A
{\bf B}_AD_{bL}\right)\nonumber\\
+\frac{2}{3}{\overline U}_{aR}\gamma^A
{\bf B}_AU_{bR}-\frac{1}{3}{\overline D}_{aR}\gamma^A
{\bf B}_AD_{bR}-{\overline E}_{aR}^-\gamma^A
{\bf B}_AE^-_{bR}]\nonumber\\
+...~~~~~~~~~~~~~~~~~~~~~~~~~~~~~~~~~~~~~~~~~~~~~~~~~~~~~~~~~~~~
\end{eqnarray}
where ${\bf V}_A^{x}$ is an SU(3) octet of gluons, ${\bf W}_A^{y}$
is an SU(2) isovector of intermediate bosons, ${\bf B}_A$
is the hypercharge boson, ${\tau_y}$ and   ${\lambda_x}$ are the 
usual Pauli and Gell-Mann matrices, and the dots 
stand for the couplings of the lepto-quark/diquark bosons to fermions.
The above result, of course, contains the SM interactions. 
Next, let us look at the  vector interaction of the 24 plet of SU(5)
in the $\overline{16}-16-210$ coupling. This can be read off from Eq.(51)
and one has
\begin{eqnarray} 
{\cal L}_{24/210}= g_{ab}^{^{(210)}} 
\left(-\overline {\bf M}_a^j\gamma^A{\bf M}_{bi}
+\frac{1}{3}\overline{\bf M}_{aik}\gamma^A{\bf M}_b^{kj}\right){\mathsf
G}_{Aj}^i\nonumber\\
=\frac{1}{3}\frac{g_{ab}^{^{(210)}}}{g_{ab}^{^{(45)}}}{\cal L}_{24/45}
-\frac{4}{3}\{g_{ab}^{^{(210)}}\sum_{x=1}^8 \overline D_{aR}\gamma^A{\bf
V}_A^{x}\frac{\lambda_x}{2}
D_{bR}\nonumber\\
+g_{ab}^{^{(210)}} 
\sqrt{\frac{3}{5}}\left[-\frac{1}{2}\left({\overline E}_{aL}^-\gamma^A
{\bf B}_AE^-_{bL}+{\overline \nu}_{aL}\gamma^A
{\bf B}_A\nu_{bL}\right)
-\frac{1}{3}{\overline D}_{aR}\gamma^A
{\bf B}_AD_{bR}\right]\nonumber\\
+g_{ab}^{^{(210)}} \sum_{y=1}^3\left(\matrix{{\overline \nu}&
{\overline E}^-}\right)_{aL}\gamma^A{\bf W}_A^{y}\frac{\tau_y}{2}
\left(\matrix{\nu\cr
E^-}\right)_{bL}+...\}
\end{eqnarray}
Eq.(67) shows that the 24 plet of SU(5) couplings in $\overline {16}-16-210$,
unlike the case of the 24 plet couplings in $\overline {16}-16-45$,
do not contain the same exact interactions as in the Standard Model.

\end{document}